\documentstyle[preprint,aps]{revtex}
\draft

\begin{document}

\title{
{\small \begin{flushright}
MPI/PhT/98-72\\[-0.2cm]
hep-ph/9809503\\[-0.2cm]
September 1998
\end{flushright} }
Neutrino oscillations in space within a solvable model}
 
\author{Ara Ioannisian\footnote[1]{On leave of absence from Yerevan
    Physics Institute, Alikhanian Br.\ 2, 375036, Yerevan, Armenia} 
and Apostolos Pilaftsis\footnote[2]{Present address:
  Theory Division, CERN, CH-1211 Geneva 23, Switzerland}}
 
\address{Max-Planck-Institut f\"ur Physik (Werner-Heisenberg-Institut),\\
F\"ohringer Ring 6, 80805 Munich, Germany}
 
\maketitle
 
\thispagestyle{empty}
 
\begin{abstract}
We study neutrino  oscillations in space within  a realistic model  in
which both  the source and the  target are considered to be stationary
having  Gaussian-form   localizations.  The   model  admits   an exact
analytic solution  in field theory which  may be expressed in terms of
complementary  error functions, thereby  allowing  for  a quantitative
discussion     of quantum-mechanical  (coherent)  versus   statistical
(incoherent) uncertainties.  The solvable model provides an insightful
framework   in  addressing   questions  related   to   propagation and
oscillation of neutrinos that may  not be  attainable by the  existing
approaches.  We find a novel  form of plane-wave behaviour of neutrino
oscillations  if the   localization spread of   the source  and target
states  due to   quantum  mechanics is  of  macroscopic  size but much
smaller  than neutrinos' oscillation length.   Finally, we discuss the
limits on the coherence length of neutrino oscillations and find that
they mainly arise  from uncertainties  of statistical origin.
\medskip


\bigskip

\centerline{\em To appear in Physical Review D}
\end{abstract}

\newpage

It was thirty  years ago when Pontecorvo  suggested \cite{BP} that  if
the   known light neutrinos  $\nu_e$,  $\nu_\mu$  and $\nu_\tau$  have
non-zero masses and mixings,  they may oscillate to  each other in the
very  much the  same  way  as the   $K^0$  and $\bar{K}^0$   mesons do
\cite{RGS}.  This  mechanism  has also been    invoked to explain  the
energy deficit  between the neutrinos  produced in the sun  or earth's
atmosphere   and those detected on earth's   ground \cite{recent}.  In
particular,  according to the recent   results of the Super-Kamiokande
collaboration \cite{exp}, the  experimental  data may  comfortably  be
explained   by naively  assuming $\nu_\mu$-to-$\nu_\tau$ oscillations,
when compared to the  non-oscillation scenario of the Standard  Model.
For neutrino  energies  $E$ much larger  than their  respective masses
$m_i$ (with $i=1,2,3$),  the classical formula \cite{BP} governing the
probability  of $\nu_\alpha$-to-$\nu_\beta$ oscillations as a function
of the distance $l$ from the source reads
\begin{equation}
  \label{Osc}
P_{\alpha\to \beta}(l)\ =\ \sum_{j=1}^3\, |V_{\alpha j}|^2 |V_{\beta j}|^2
\ +\ 2\sum_{j>k}^3\, \Re e (  V_{\alpha j} V_{\beta j}^*
V_{\alpha k}^* V_{\beta k})\, \cos \Big( \frac{(m^2_j - m^2_k) l}{2E}\,
\Big)\ ,
\end{equation}
where  $V_{\alpha j}$  is a  leptonic  mixing matrix analogous  to the
Cabbibo-Kobayashi-Maskawa  mixing matrix  of   the quark sector.   For
simplicity, we have also neglected possible effects of CP violation in
Eq.\      (\ref{Osc}).       Up       to   now,      many      authors
\cite{SN/BK,GKL,Rich,LBBRS,DMOS,KNW,GS,JEC,HJL,KW,YVS,LES,GK}     have
discussed the conditions under which the above naive cosine formula of
oscillations between  neutrinos or other particles  may or may  not be
valid.  In a large number of papers \cite{SN/BK,GKL,LBBRS,KNW,HJL,GK},
the derivation of  Eq.\ (\ref{Osc}) has been  based on  using neutrino
wave  packets.     Within  other  more    field-theoretically oriented
approaches \cite{Rich,GS,JEC,KW},  neutrinos  are  treated as  virtual
intermediate states which are produced by  some initial states and are
only observed  through their interaction with  the target state in the
detector. The  latter considerations may also be   related to those of
the old-fashioned S-matrix  approach  \cite{RGS,DI}.  However, in  all
different  treatments present in   the literature, many approximations
were   necessary to  arrive at   the    known formula (\ref{Osc})   of
oscillations.

In this paper, we shall study  particle oscillations in space within a
model in   which both the  source  and target states   are taken to be
stationary having Gaussian-type broadenings. For our illustrations, we
first consider the oscillation of  two scalar particles whose dynamics
may well be determined by the relevant propagators at the second order
of electroweak interactions.  The  discussion is then extended  to the
case of  fermions, {\em i.e.}, that of  neutrinos.  Since  we assume a
Gaussian-form   localization  for   the   production   and   detection
interactions, the model admits  an exact analytic solution within  the
framework of field theory  and hence different sorts of approximations
can  directly  be controlled for  their  validity.  Thus, the solvable
model offers an important insight into  the profound issue of particle
oscillations, thereby  complementing  related  approaches discussed in
\cite{Rich,GS,JEC,KW}.   In particular, we  find  that the oscillating
pattern of neutrinos depends  crucially on the coherent  broadening of
production and detection  points.  As long  as the coherent broadening
effects lie within the microscopic range, one has the usual picture of
particle propagation  through spherical waves in the three-dimensional
space, as the particles are emitted from a well-localized source, {\em
  e.g.},   neutrinos from the sun.     If such a broadening,  however,
happens to be of mesoscopic size {\em but still}  is much smaller than
the oscillation length of  particles, then their  propagation proceeds
via plane waves and their oscillation length depends explicitly on the
three momentum direction of the initial states at the source.

Let  us  consider the  $2\to   2$ scattering process  with  two scalar
particles as  intermediate states which may  oscillate to one another. 
The  production and  detection of   the  intermediate scalars   occurs
through some asymptotic   states  at the  space   points $\bar{x}$ and
$\bar{y}$, respectively.  For  thermal-source  situations under  study
\cite{LES}, the time  elapsed between production and  detection cannot
be measured directly. Therefore, it is  legitimate to assume that to a
rather good approximation, the interactions taking place in the source
and target are stationary.  The two points $\bar{x}$ and $\bar{y}$ are
then   macroscopically separated   by a  distance   $l   = |\bar{x}  -
\bar{y}|$.  For example,  for  solar neutrinos   $l$ is the   distance
between sun  and earth, for  atmospheric neutrinos $l$ is the distance
between earth's atmosphere   and detector at  the ground,   {\em etc.} 
Furthermore, we take the spatial localizations  of the interactions at
the source and target to be Gaussian functions peaked at $\bar{x}$ and
$\bar{y}$ with  dispersions of cubic form, {\em   i.e.}, $\delta x^i =
\delta x$ and $\delta y^i = \delta y$.

With the above considerations,  the amplitude  for $\alpha \to  \beta$
transition is given by
\begin{eqnarray}
  \label{Tamp1} 
{\cal T}_{\beta\alpha} (k,\bar{y},\delta y ; p,\bar{x},\delta x) &=& 
\widetilde{\cal N}\,
\int d^4x\, d^4y\ e^{-(\vec{x}-\bar{x})^2/\delta x^2}\ 
e^{-(\vec{y}-\bar{y})^2/\delta y^2}\ e^{-ipx + iky}\nonumber\\
&\times& \sum_{j=1}^2\ 
O^{\alpha\beta}_j\ \int \frac{d^4q}{(2\pi)^4}\ \frac{e^{iq(x-y)}}{
q^2 - m^2_j + i\varepsilon}\ ,
\end{eqnarray}
where $p = p_1 + p_2$ and $k = k_1 + k_2$ are the total momenta of the
initial  and  final states, respectively. The  overall proportionality
factor $\widetilde{\cal  N}$  contains  coupling constants  and  other
irrelevant  multiplicative terms; it  drops  out after the probability
rate is normalized. Furthermore,
\begin{equation}
  \label{Cabj}
O^{\alpha\beta}_j\ =\ O_{\alpha j} O_{\beta j}\, 
\end{equation}
is the  usual combination of elements of  the mixing  matrix $O$.  The
$2\times 2$ orthogonal matrix $O$ diagonalizes the  mass matrix of the
two-scalar  system, and   relates the weak  to  the  mass eigenstates.
Obviously, $\sum_j   O^{\alpha\beta}_j = \delta_{\alpha\beta}$.  Here,
we should  also remark   that   Lorentz invariance of the    amplitude
(\ref{Tamp1}) is not  manifest; it  is  broken by the  space-dependent
Gaussian functions.  However, one can always recover the manifest form
of  Lorentz  invariance by  considering  instead the Lorentz-invariant
exponent $\delta k_\mu  (x -\bar{x})^\mu$  for the Gaussian   function
with $\delta k_\mu = ( 1/\delta x^0, 1/\delta x^i  )$ and likewise for
the $y$ coordinate.  The stationarity  condition implies that  $\delta
x^0 \gg \delta x^i$ and $\delta y^0 \gg \delta y^i$.  For our inertial
frame of particle production   and  detection, we take $\delta   x^0,\
\delta y^0 \to \infty$.

Since the interactions at $\bar{x}$  and $\bar{y}$ do not display  any
time  dependence  in  the   stationary  limit  discussed   above,  the
integration   over  the times  $x^0$   and $y^0$   can  now easily  be
performed. The  transition  amplitude  ${\cal  T}_{\beta\alpha}$  then
takes the form
\begin{eqnarray}
  \label{Tamp2} 
{\cal T}_{\beta\alpha} (k,\bar{y},\delta y ; p,\bar{x},\delta x ) &=& 
\widetilde{\cal N}\, \delta (k^0-p^0)\ \int d^3\vec{x}\, d^3\vec{y}\ 
e^{-(\vec{x}-\bar{x})^2/\delta x^2}\ 
e^{-(\vec{y}-\bar{y})^2/\delta y^2}\ e^{i\vec{p}\vec{x} - i\vec{k}\vec{y}} 
                                           \nonumber\\
&\times& \sum_{j=1}^2\ O^{\alpha\beta}_j\ \int \frac{d^3\vec{q}}{(2\pi)^3}\ 
\frac{e^{-i\vec{q}(\vec{x} - \vec{y})}}{q^2_j - |\vec{q}|^2 + i\varepsilon}\ ,
\end{eqnarray}
where $q^2_j  = E^2 - m^2_j$. The  result of the time  integrations is
that   energy  conservation holds   strictly  both  at production  and
detection vertices,  {\em i.e.}, $k^0=p^0=E$,  as  can readily be seen
from the delta function on the RHS of Eq.\ (\ref{Tamp2}).

It  is now important to  notice that under spatial displacements, {\em
e.g.},  $\bar{x}\to \bar{x}   +  \vec{a}$ and  $\bar{y}\to  \bar{y}  +
\vec{a}$, ${\cal  T}_{\beta\alpha}  ( \bar{x}, \bar{y}   )$ transforms
into $e^{i\vec{a}(\vec{k}    -   \vec{p})} {\cal   T}_{\beta\alpha}  (
\bar{x},  \bar{y}    )$.   In    other    words,  the   amplitude   is
frame-independent up    to an  unobservable   phase $\vec{a}(\vec{k} -
\vec{p})$.  Employing this fact, it  proves convenient to redefine the
amplitude into a manifestly frame-independent form
\begin{eqnarray}
  \label{Tamp3} 
{\cal T}_{\beta\alpha} (k,p,\vec{l},\delta y, \delta x) &\to& 
e^{-i\vec{p}\bar{x} + i\vec{k} \bar{y}}\ 
{\cal T}_{\beta\alpha} (k,\bar{y},\delta y ; p,\bar{x},\delta x)
\nonumber\\
&=& \widetilde{\cal N}\,
\delta (k^0-p^0)\ \int d^3\vec{x}\, d^3\vec{y}\ 
e^{-\vec{x}^2/\delta x^2}\ 
e^{-\vec{y}^2/\delta y^2}\ e^{i\vec{p}\vec{x} - i\vec{k}\vec{y}} 
                                           \nonumber\\
&\times& \sum_{j=1}^2\ O^{\alpha\beta}_j\ \int \frac{d^3\vec{q}}{(2\pi)^3}\ 
\frac{e^{- i\vec{q}(\vec{x} - \vec{y} - \vec{l})}}{q^2_j - |\vec{q}|^2 
+ i\varepsilon}\ , 
\end{eqnarray}
with $\vec{l} = \bar{y}-\bar{x}$.   It is now  not difficult  to carry
out the Gaussian integrals over $\vec{x}$ and $\vec{y}$ (see also Eq.\
(\ref{Gauss}) in Appendix A). In this way, we obtain
\begin{eqnarray}
  \label{Tamp4} 
{\cal T}_{\beta\alpha} (k,p,\vec{l},\delta y, \delta x)\ =\ {\cal N}\,
\delta (k^0-p^0)\ \sum_{j=1}^2\ O^{\alpha\beta}_j\ 
\int \frac{d^3\vec{q}}{(2\pi)^3}\  
\frac{e^{-\delta x^2 (\vec{p} - \vec{q})^2/4}\ 
e^{-\delta y^2 (\vec{k} - \vec{q})^2/4}\
e^{i\vec{q}\vec{l}} }{q^2_j - |\vec{q}|^2 + i\varepsilon}\ ,
\end{eqnarray}
with ${\cal N}  = \pi^3 \delta  x^3  \delta y^3\, \widetilde{N}$.   In
Appendix A we show that the very  last three-dimensional integral over
$\vec{q}$ in  Eq.\ (\ref{Tamp4}) can be  solved exactly within a class
of error functions.  Taking this very last fact into consideration, we
arrive at the following analytic expression:
\begin{eqnarray}
  \label{Tamp5} 
{\cal T}_{\beta\alpha} (k,p,\vec{l},\delta y, \delta x) &=& -\, {\cal N}\,
     \delta (k^0-p^0)\ \frac{1}{8\pi |\vec{L}| }\ 
     \sum_{j=1}^2\ O^{\alpha\beta}_j\  
     e^{-\frac{1}{4}\,\delta x^2 (|\vec{p}|^2+q^2_j)  - 
     \frac{1}{4}\, \delta y^2 (|\vec{k}|^2 +q^2_j)}\nonumber\\
&& \hspace{-2cm} \times\, \Big[\, 
e^{iq_j |\vec{L}|}\, {\rm Erfc} \Big(-\frac{i}{2}\delta l q_j
-\frac{|\vec{L}|}{\delta l}\Big)\ -\ e^{-iq_j |\vec{L}| }\, 
{\rm Erfc} \Big( -\frac{i}{2}\delta l q_j +\frac{|\vec{L}|}{\delta l}\Big)\, 
\Big]\ ,
\end{eqnarray}
where  Erfc$(z)$ is the complementary  error  function defined in  the
appendix, $\delta l^2 = \delta x^2 + \delta y^2$ and
\begin{equation}
  \label{Lvec}
\vec{L}\ =\ \vec{l}\, -\, \frac{i}{2}\,\delta x^2 \vec{p}\, -\, 
\frac{i}{2}\,\delta y^2 \vec{k}\, .
\end{equation}
The norm of the complex vector $\vec{L}$ is  understood to act only on
the vectorial space,  {\em i.e.}, $|\vec{L}| \equiv \sqrt{\vec{L}^2}$.
Equation (\ref{Tamp5})  is the major result of  this paper, from which
any  behaviour  of neutrino    oscillations in  space    for different
kinematic conditions may directly be inferred.

On  physical grounds,  one  generally expects  that $\vec{k}-\vec{p} =
{\cal O}(1/\delta x, 1/\delta y)$.  Therefore, without sacrificing any
of the physical features of particle oscillations we wish to study, we
can explicitly impose  a kind of  three-momentum  conservation for the
initial and final states of the scattering in Eq.\ (\ref{Tamp5}), {\em
  i.e.},  $\vec{k}   =  \vec{p}$   \cite{DI,DMOS}.  This   leaves  the
three-momentum  of the  oscillating  particles still unspecified.   In
fact, the very same result would have been obtained, if we had started
with a   Gaussian  interaction of the   form  $\exp [-(\vec{x}-\vec{y}
-\vec{l})^2/ \delta l^2]$, with $\delta l^2 = \delta x^2 + \delta y^2$
and integrated out  first the  frame-dependent components $(\vec{x}  +
\vec{y})/2$. However, we should stress  that our conclusions would not
change even if we considered  the complete but more lengthy expression
(\ref{Tamp5}).   Under   these assumptions,  the probability amplitude
simplifies to
\begin{eqnarray}
  \label{TampS} 
{\cal T}_{\beta\alpha} (k,\vec{l},\delta l) &=& -\, {\cal N}\,
\delta^{(4)} (k - p)\ \frac{1}{ 8\pi\,
|\vec{l}-\frac{i}{2}\delta l^2 \vec{k}| }\
\sum_{j=1}^2\ O^{\alpha\beta}_j\ 
 e^{-\frac{1}{4}\,\delta l^2 (|\vec{k}|^2+q^2_j)} \ \Big[\,
e^{iq_j |\vec{l} - \frac{i}{2}\delta l^2 \vec{k} | }\nonumber\\ 
&&\hspace{-2.cm}\times {\rm Erfc}\Big(-\frac{i}{2}\delta l q_j -
\frac{|\vec{l} - \frac{i}{2}\delta l^2 \vec{k}|}{\delta l}\Big)\ -\
e^{-iq_j |\vec{l} - \frac{i}{2}\delta l^2 \vec{k}|}\,
{\rm Erfc}\Big(-\frac{i}{2}\delta l q_j + \frac{|\vec{l} 
- \frac{i}{2}\delta l^2 \vec{k}|}{\delta l}\Big)\, \Big]\, . 
\end{eqnarray}
Notice  that ${\cal N}$ must  be  redefined accordingly for reasons of
dimensionality.  We  can now determine the  probability for  the state
$\alpha$ produced at  a  distant well-localized area, {\em  e.g.}, the
sun, to observe the state $\beta$ at  a macroscopic distance $l$.  The
normalized probability is given by
\begin{equation}
  \label{Pab}
\frac{d}{d\Gamma}\ P_{\alpha\to \beta}(k,l,\delta l)\ =\
\frac{ |{\cal T}_{\beta\alpha} (k,l,\delta l)|^2 }
{\sum_\beta\  \int d\Gamma\ |{\cal T}_{\beta\alpha} (k,l,\delta l)|^2}\ ,
\end{equation}
where $\Gamma$ denotes  the phase space of the  final states.  In Eq.\
(\ref{Pab}), we have  not  yet included uncertainties  of  statistical
origin which can be added incoherently (we defer the discussion to the
end of  the paper). In the following,  we shall focus our attention on
coherent quantum-mechanical uncertainties.

It is very instructive to  discuss  the following two limiting  cases:
(i) $\delta l^2  |\vec{k}| \ll l$ and (ii)  $\delta  l^2 |\vec{k}| \gg
l$. Let us first  consider case (i).  We  can expand  the norm of  the
complex vector   in   Eq.\  (\ref{TampS})  in  terms of   $\delta  l^2
|\vec{k}|$, {\em i.e.}
\begin{equation}
  \label{case1}
|\vec{l} - \frac{i}{2}\delta l^2 \vec{k} |\ =\ l \, -\, 
    \frac{i}{2}\,\delta l^2\, \frac{\vec{k}\vec{l}}{l}\ +\ 
        {\cal O}\Big( \frac{\delta l^4 |\vec{k}|^2}{l^2}\,\Big)\ .
\end{equation}
To  leading order, the  transition amplitude then   takes on the known
oscillatory form
\begin{equation}
  \label{TampS1} 
{\cal T}_{\beta\alpha} (k,\vec{l},\delta l) \ =\ -\, {\cal N}\,
\delta^{(4)} (k - p)\ \frac{1}{4\pi\, l}\
\sum_{j=1}^2\ O^{\alpha\beta}_j\ 
e^{-\frac{1}{4}\,\delta l^2 (\vec{k} - q_j\vec{l}/l )^2}\ 
e^{iq_j l}\ +\ {\cal O}\Big( \frac{\delta l^4 |\vec{k}|^2}{l^2}\,\Big)\ .
\end{equation}
The   $\vec{k}$-dependent exponential  factor in  Eq.\  (\ref{TampS1})
controls the flow of the  three-momentum between the asymptotic states
and the  intermediate particles,  {\em i.e.},  that of  neutrinos.  In
fact, the three-momentum is conserved  in the production and detection
vertices up to an  error of order $1/\delta l$.   In the limit $\delta
l\to  0$, any information about  the three-momentum of the oscillating
system  is completely  lost,   {\em  e.g.}, the  total  three-momentum
$|\vec{k}|$ of the  detecting particles can  take any possible  value. 
However,  in experiments  the   distribution  over $\vec{k}$  can   in
principle be measured, so physically  $\delta l$ may  be small but not
zero. How small  or how big $\delta l$   could be is a  puzzling issue
that will be discussed below.

After integrating over the three-momentum phase space $\Gamma$ in Eq.\
(\ref{Pab}), we find the well-known formula of particle oscillations
\begin{equation}
  \label{Posc}
P_{\alpha\to \beta}(E,l,\delta l)\ =\ \Big|\sum_{j=1}^2\ 
O^{\alpha\beta}_j\, e^{iq_j l}\, \Big|^2\ +\ 
{\cal O} \Big( \delta l (q_1-q_2), \frac{\delta l^2 q_j}{l}\,  \Big)\ . 
\end{equation}
The above formula is valid under the following two conditions:
\begin{eqnarray}
  \label{cond1}
\mbox{I.}&&\quad \delta l\ \ll\ L^{\rm osc}\ =\ \frac{2\pi}{q_1-q_2}\ ,\\
  \label{cond2}
\mbox{II.}&&\quad \delta l^2 q_j\ \ll\ l\ .
\end{eqnarray}
The first requirement reflects the  fact that the coherent uncertainty
in the position $\delta  l$ for the source  and target states  must be
much smaller  than the  oscillation length  $L^{\rm osc}$ in  order to
have a non-vanishing  oscillating pattern.  Within the field-theoretic
framework,     this constraint  has   first been   discussed  in Ref.\ 
\cite{Rich}  and  further elaborated on in  \cite{GS,JEC,KW,YVS}. Note
that a constraint analogous to Eq.\ (\ref{cond1}) may also be obtained
in the wave-packet treatment if $\delta l$ is to be interpreted as the
size of  the  neutrino  wave packets  \cite{SN/BK,GKL,KNW,HJL,LES,GK}. 
Nevertheless,  the new requirement  in Eq.\ (\ref{cond2}) also affects
the form of particle oscillations and deserves high attention as well.
In  particular,  it is  interesting to  notice  that  one can  satisfy
condition I  by grossly violating condition II.   To  give an example,
suppose that $\delta l = 1$ cm, $\Delta m^2 = m^2_1 - m^2_2 = 10^{-4}$
eV$^2$, and the energy $E  = 1$ GeV (1  GeV $\approx 5 \times 10^{13}\ 
{\rm cm}^{-1}$) which is typical for atmospheric neutrinos.  Then, one
has $L^{\rm osc} \approx E/\Delta m^2 \approx 10^8$ cm, certainly much
larger than $\delta l$.  On the other hand, one finds that $\delta l^2
q_1 \approx  (1\ {\rm  cm})^2  \times 1\  {\rm GeV}  \approx  5 \times
10^{13}$ cm is comparable to the distance between sun and earth!

To elucidate  further the  implications of the  new  condition  II for
particle oscillations, we shall    now discuss the behaviour   of  the
probability amplitude for  the   limiting case (ii) when  $\delta  l^2
|\vec{k}| \gg  l$. In this   limit, the  norm   of the complex  vector
$\vec{l} - \frac{i}{2}\delta l^2 \vec{k}$ may be expanded as follows:
\begin{equation}
  \label{case2}
|\vec{l} - \frac{i}{2}\delta l^2 \vec{k} |\ =\  - \ 
\frac{i}{2}\, \delta l^2 |\vec{k}|\ +\ \frac{\vec{k} \vec{l}}{|\vec{k}|} \ +\ 
        {\cal O}\Big( \frac{ l^2 }{\delta l^4 |\vec{k}|^2} \,\Big)\ .
\end{equation}
Substituting the last expression into Eq.\ (\ref{TampS}) and employing
the asymptotic expansion of ${\rm Erfc}(z)$ for large values of $z$ in
Eq.\ (\ref{Erfcexp}) yields
\begin{equation}
  \label{TampS2} 
{\cal T}_{\beta\alpha} (k,\vec{l},\delta l) \ =\ -{\cal N}\,
\delta^{(4)} (k - p)\ \frac{i}{2\pi \delta l^2 |\vec{k}|}\
\sum_{j=1}^2\ O^{\alpha\beta}_j\ 
e^{-\frac{1}{4}\,\delta l^2 (|\vec{k}| - q_j )^2}\ 
e^{iq_j \vec{k}\vec{l}/|\vec{k}| }\ +\ 
{\cal O}\Big( \frac{l^2}{ \delta l^4 |\vec{k}|^2 }\,\Big)\ ,
\end{equation}
Because    of  the absence     of  the  $1/l$   dependence of   ${\cal
  T}_{\beta\alpha}$  in Eq.\  (\ref{TampS2}),  the propagation of  the
intermediate states turns over to  plane waves depending explicitly on
the three momentum $\vec{k}$.  This intriguing plane-wave behaviour of
the amplitude may  be  attributed to  the fact that   non-localization
effects of the source and target quantum  states within a finite space
volume \cite{GCH}  get  coherently amplified   for large  (mesoscopic)
values  of  $\delta l$.   Geometrically, one  may attempt to visualize
this counter-intuitive result of the behaviour of particle propagation
as follows.   It  appears that   the intermediate states   are emitted
and/or detected by  means of an  `antenna' of macroscopic size $\delta
l^2  |\vec{k}|$ instead of  $\delta  l$ that   one would have  naively
expected.  So, if an observer were close to such an `antenna' emitting
neutrinos, say, in a distance  much smaller to  its size, she/he would
find in principle  that these particles  are  emitted by  plane waves,
whereas she/he   would  only recover   the  usual form of  propagation
through spherical waves if she/he were  at a distance much bigger than
the size of the `antenna.'  Nevertheless,  we still feel that a deeper
understanding of  this  quantum-mechanical  phenomenon would  be  very
useful.

One  might now  worry that averaging  effects  due to  the phase-space
integration  of the final states  would cancel the factor proportional
to $\vec{k}\vec{l}$ which appears    in the oscillating part   of  the
probability in Eq.\ (\ref{TampS2}).  However, this is not  quite true,
since  the square of  the transition amplitude ${\cal T}_{\beta\alpha}
(k,\vec{l},\delta l)$ does not depend on the three-momentum difference
$k_- = k_1 - k_2$, and hence the phase-space integration over $\Gamma$
has   the  effect   to   replace    simply  the  oscillatory    factor
$\vec{k}\vec{l}$ with   $\vec{p}\vec{l}$.  To  make this  explicit, we
note that one can always make  the following variable substitutions in
the phase-space integral:
\begin{eqnarray}
  \label{LIPS}
\int d^4k_1\, d^4k_2\,
\delta_+ (k^2_1-m^2_1)\, \delta_+ (k^2_2 - m^2_2)\, \delta^{(4)}(k-p)\,
|{\cal T}_{\beta\alpha}(k,\vec{l},\delta l)|^2 &&\nonumber\\
 &&\hspace{-10.5cm}=\ 
\frac{1}{2}\, \int d^4k\, d^4k_-\,
\delta_+ \Big[ \frac{(k+k_-)^2}{4} - m^2_1\Big]\, 
\delta_+ \Big[\frac{(k-k_-)^2}{4} - m^2_2 \Big]\, 
\delta^{(4)}(k-p)\, |{\cal T}_{\beta\alpha}(k,\vec{l},\delta l)|^2\nonumber\\
&&\hspace{-10.5cm}=\ \frac{1}{2}\, 
|{\cal T}_{\beta\alpha}(p,\vec{l},\delta l)|^2\, 
\int d^4k_-\, \delta_+ \Big[\frac{(p+k_-)^2}{4} - m^2_1\Big]\, 
\delta_+ \Big[ \frac{(p-k_-)^2}{4} - m^2_2\Big]\, ,
\end{eqnarray}
where $\delta_+  (k^2_i  - m^2_i) =  \theta   (k^0_i) \delta (k^2_i  -
m^2_i)$  ($i  =  1,2$).   The  integral in   the  last   step of  Eq.\
(\ref{LIPS}) is just an overall  normalization constant and cancels in
Eq.\  (\ref{Pab}).  Taking  this   into  account,  the probability  of
particle oscillations reads
\begin{equation}
  \label{PoscII}
P_{\alpha\to \beta} (E,\vec{p},\vec{l},\delta l)\ =\ \Big|\sum_{j=1}^2\ 
O^{\alpha\beta}_j\, e^{iq_j \vec{p}\vec{l}/|\vec{p}|}\, \Big|^2\ +\ 
{\cal O} \Big( \delta l (q_1 - q_2), \frac{l}{\delta l^2 q_j}\,  \Big)\ . 
\end{equation}
{}From the above formula, it is obvious that the oscillation length as
seen by  the detector is strongly   correlated with the three-momentum
$\vec{p}$ of the  initial states producing  the  mixed particles.   In
principle, the two  different production mechanisms  as well  as their
predictions obtained by Eqs.\ (\ref{Posc}) and (\ref{PoscII}) could be
distinguished if we were able to measure the spatial dependence of the
oscillation length   by moving the   detector around the   source.  In
addition, we must  assume that one  somehow knows the initial momentum
$\vec{p}$ of the production mechanism  and controls well both coherent
($\delta y$) and incoherent uncertainties at the detector. Even though
this seems to be a rather formidable  task, the actual size of $\delta
l$ dictates the   nature of particle  propagation within  the solvable
model.  So   far, many  qualitative estimates  exist   for $\delta l$. 
However, there has not been  yet  a rigorous  method to evaluate  this
quantity from first principles.  It is obvious that $\delta l$ depends
decisively  on   the  details   of  the   experiment.     According to
\cite{SN/BK,HJL,KNW,LES},  the  most natural choice  for  the scale of
$\delta l$ for neutrinos coming from the sun lies  between the size of
nucleus,  {\em  i.e.}\ several   fm's, and  $10^{-7}$  cm.  The latter
originates from  effects due to  thermal collisions of electrons which
results  in a  reduction of  the  wave-packet  size of the  production
system.  Therefore, it  is most likely to  expect that solar neutrinos
have a spherical-wave propagation.

We shall now turn to the case of oscillations  of two fermions such as
neutrinos.  Our starting point  is a transition amplitude analogous to
Eq.\ (\ref{Tamp3})
\begin{eqnarray}
  \label{Tampf1}
{\cal T}_{\beta\alpha} (k,p,\vec{l},\delta y, \delta x ) & = &  
\widetilde{\cal N}\, \delta (k^0-p^0)\ J^\mu_f (k_1) \bar{u}_\beta (k_2)\,
\gamma_\mu\, P_L\ \int d^3\vec{x}\, d^3\vec{y}\ 
e^{-\vec{x}^2/\delta x^2}\ 
e^{-\vec{y}^2/\delta y^2}\ e^{i\vec{p}\vec{x} - i\vec{k}\vec{y}} 
                                           \nonumber\\
&\times& \sum_{j=1}^2\ V^{\alpha\beta}_j\ \int \frac{d^3\vec{q}}{(2\pi)^3}\ 
\frac{(\gamma^0 E - \vec{\gamma} \vec{q} + m_j )\ 
e^{- i\vec{q}(\vec{x} - \vec{y} - \vec{l})}}{
q^2_j - |\vec{q}|^2 + i\varepsilon}\ \gamma_\nu P_L 
u_\alpha (p_2) J^\nu_i (p_1)\nonumber\\ 
&=& {\cal N}\, \delta (k^0-p^0)\ J^\mu_f (k_1) \bar{u}_\beta (k_2)\,
\gamma_\mu\, P_L\ \nonumber\\
&\times& \sum_{j=1}^2\ V^{\alpha\beta}_j\ 
( \gamma^0 E\, -\, \vec{\gamma}\, \nabla_{\vec{l}} )\
\int \frac{d^3\vec{q}}{(2\pi)^3}\  
\frac{e^{-\delta x^2 (\vec{p} - \vec{q})^2/4}\ 
e^{-\delta y^2 (\vec{k} - \vec{q})^2/4}\ 
e^{i\vec{q}\vec{l}} }{q^2_j - |\vec{q}|^2 + i\varepsilon}\nonumber\\ 
&\times& \gamma_\nu u_\alpha (p_2) J^\nu_i (p_1)\ .
\end{eqnarray}
where $p = p_1 + p_2$, $k = k_1 + k_2$, $P_L = (1 - \gamma_5)/2$, $E =
k^0 = p^0$ ($q_j=\sqrt{E^2 - m^2_j}$) and
\begin{equation}
  \label{Vabj}
V^{\alpha\beta}_j\ =\ V_{\alpha j} V^*_{\beta j}\, .
\end{equation}
Specifically,  Eq.\ (\ref{Tampf1}) describes the amplitude probability
for producing the neutrino flavour $\alpha$ at point $\vec{x}$ through
the standard   $V-A$  weak  current  interaction  $J^\nu_i  (p_1)$ and
detecting  the neutrino flavour $\beta$  through  a similar $V-A$ weak
current  interaction $J^\mu_f (k_1)$.    The neutrino   flavour states
$\alpha$ and  $\beta$  may   be identified  by  the   charged  leptons
$l_\alpha$   and $l_\beta$  which     accompany the neutrinos  in  the
production and detection vertices, respectively.

The analytic  form of  the  transition amplitude  describing  neutrino
oscillations is given by 
\begin{eqnarray}
  \label{TampF} 
{\cal T}_{\beta\alpha} (k,p,\vec{l},\delta y,\delta x) &=& -\, {\cal N}\,
\delta (k^0 - p^0)\ \frac{1}{ 8\pi\, |\vec{L}| }\
\sum_{j=1}^2\ V^{\alpha\beta}_j\ 
 e^{-\frac{1}{4}\,\delta x^2 (|\vec{p}|^2+q^2_j)
-\frac{1}{4}\,\delta y^2 (|\vec{k}|^2+q^2_j) }\nonumber\\
&&\hspace{-3cm}\times\, \bigg( 
J^\mu_f (k_1) \bar{u}_\beta (k_2)\,
\gamma_\mu\, P_L\, E\gamma^0\, \gamma_\nu u_\alpha (p_2) J^\nu_i (p_1)\
\Big( e^{iq_j |\vec{L}|} {\rm Erfc}z^{(j)}_-\, - \, 
e^{-iq_j |\vec{L}|} {\rm Erfc}z^{(j)}_+ \Big)\nonumber\\
&&\hspace{-3cm} -\, J^\mu_f (k_1) \bar{u}_\beta (k_2)\,
\gamma_\mu\, P_L\, \frac{\vec{\gamma}\,\vec{L} }{|\vec{L}|}\,
\gamma_\nu u_\alpha (p_2) J^\nu_i (p_1)\
\Big\{ e^{iq_j |\vec{L}|}\, \Big[ \Big( q_j - 
\frac{1}{|\vec{L}|} \Big)\, {\rm Erfc}z^{(j)}_-\, -\, \frac{2}{\sqrt{\pi} 
\delta l}\, e^{-z^{(j)2}_-}\, \Big] \nonumber\\
&& -\, e^{-iq_j |\vec{L}|}\, \Big[ \Big( q_j + 
\frac{1}{|\vec{L}|} \Big)\, {\rm Erfc}z^{(j)}_+\, -\, \frac{2}{\sqrt{\pi} 
\delta l}\, e^{-z^{(j)2}_+}\, \Big]\, \Big\}\, \bigg),
\end{eqnarray}
where the complex vector $\vec{L}$ is defined in Eq.\ (\ref{Lvec}) and
$z^{(j)}_\mp = -\frac{i}{2}\delta l^2 q_j\, \mp\, |\vec{L}|/\delta l$.
Apart from obvious complications due to the spinorial structure of the
analytic expression (\ref{TampF}), the theoretical predictions as well
as conclusions concerning  neutrino oscillations are very analogous to
the scalar case discussed  above, and we shall  not repeat these here.
Instead, we wish to make a few clarifying remarks concerning the model
under discussion.  Even  though the Gaussian-type spatial  broadenings
are  equivalent  to     considering wave-packets  for    the   initial
(production)   and  final      (detection)  system   of   states,  our
field-theoretic approach does not  involve explicitly wave-packets for
the propagating    particles.   In   fact,  the propagation    of  the
quasi-virtual intermediate states in  space is consistently taken into
account  by  their  respective  propagators without  recourse  to  any
boundary or preparation condition  for the propagating  states.  Since
we integrate  over    all times,   energy conservation    and  flavour
factorization at the source is a consequence  of our model rather than
an input to it.

Because of  the strict  energy  conservation discussed  above  both at
production and   detection     vertices, the formulas    of   particle
oscillations in   Eqs.\  (\ref{Posc})  and (\ref{PoscII})   predict an
infinitely  long  oscillating  pattern   \cite{GS}.  This is   not  an
unexpected consequence of field theory.  In relativistic quantum field
theory,   energy   and  three-momentum  are   described   by commuting
independent operators in the same way as time and space variables are.
This fact is also  manifested  in  the Fock-Schwinger formulation   of
propagators  \cite{FS}.  As a consequence,   energy  and space may  in
principle be  measured to  infinite precision,  despite the spread  in
three-momentum and time for the source and target states. In fact, the
solvable model under discussion amounts to assuming an infinite spread
in  time  which in turn is  equivalent  to an infinite  sharply peaked
energy distribution.     Within this framework,   the use  of on-shell
kinematic conditions to predict the energy of  a quantum system by its
three momentum is not  a rather well  justified procedure.  As we will
see below, an upper limit on the length of particle oscillations comes
from statistical (incoherent) uncertainties within the solvable model.

In addition to the coherent quantum-mechanical uncertainties discussed
above, one  has to worry about  statistical effects due  to the energy
spread  of  the stationary  source  or due to source's  and detector's
finite sizes in space  which may give rise  to destructive  effects on
particle oscillations.  Let us  discuss an illustrative example of the
kind.  Suppose that the validity conditions  for oscillations in Eqs.\
(\ref{cond1}) and (\ref{cond2}) are satisfied and  the source being in
a stationary situation emits neutrinos  of an average energy $\bar{E}$
with dispersion $\Delta E$, described by the function
\begin{equation}   
  \label{gate}
\Pi (E;\bar{E},\Delta E)\ =\ \frac{1}{\Delta E}\ \theta \Big(\, 
\frac{1}{2}\Delta E - |E-\bar{E}|\, \Big)\ .
\end{equation}
The function $\Pi   (E)$ has the shape  of  a double-well with   width
$\Delta E$ and  goes over to a delta-function  in the limit  $\Delta E
\to 0$.  Then, for non-zero values of $\delta E$, the oscillating part
of the averaged probability for $\alpha\to  \beta$ transition is given
by
\begin{eqnarray}
  \label{PEfin}
{\cal P}^{\rm osc}_{\alpha \to \beta} (l) & =& 
\int_{-\infty}^{+\infty} dE\  \Pi (E;\bar{E},\Delta E)\ 
P^{\rm osc}_{\alpha \to \beta} (E,l) \nonumber\\ 
&\approx&
2\, \Re e (V^{ab}_1 V^{ab*}_2)\ 
\frac{1}{\Delta  E}\ \int_{-\Delta  E/2}^  {\Delta E/2}\,  
dx\ \cos\,\Big[\, \frac{\Delta  m^2\,  l}{2\bar{E}}\ \Big(1  -
\frac{x}{\bar{E}}\Big)\,    \Big]\nonumber\\   
&=& 
2\, \Re e ( V^{ab}_1 V^{ab*}_2)\ \frac{\bar{E}}{\Delta E}\, 
\frac{L^{\rm osc}(\bar{E})}{\pi l}\, 
\sin 
\Big(\,\frac{\Delta E}{\bar{E}}\, \frac{\pi l}{L^{osc} (\bar{E})}\, \Big)\,
 \cos  \Big(\,\frac{2\pi l}{L^{osc} (\bar{E})}\, \Big)\ ,
\end{eqnarray}  
where we have  implicitly  assumed invariance  of our model  under CP,
{\em  i.e.},  $\Im  m (V^{ab}_1  V^{ab*}_2)   = 0$.  {}From  the  last
expression in Eq.\ (\ref{PEfin}), we readily  see that neutrinos cease
to oscillate after travelling a distance
\begin{equation}
  \label{Lcoh}
l\ >\  L^{\rm coh}\ =\ \frac{\bar{E}}{\pi \Delta E}\ L^{\rm osc}(\bar{E})\ . 
\end{equation}
In   the   literature $L^{\rm coh}$ is     called  coherence length of
oscillations, which characterizes  the maximal scale at which neutrino
oscillations can still be observed (see,  {\em e.g.}, \cite{KNW}).  By
analogy,  one can  estimate  averaging effects  on oscillations due to
source's  and  detector's  finite  sizes $\sigma_x$  and   $\sigma_y$,
respectively.  Going  through a similar   exercise, one finds for  the
oscillating part of the probability
\begin{equation}
  \label{sigmal}
{\cal P}^{\rm osc}_{\alpha \to \beta} (l)\ =\ 2\, \Re e ( V^{ab}_1 V^{ab*}_2)\ 
\frac{ L^{\rm osc}}{\pi \sigma_l}\, \sin \Big(\, \frac{2\pi \sigma_l}{
L^{\rm osc}}\, \Big)\ \cos \Big(\, \frac{2\pi l}{L^{\rm osc}}\, \Big)\ ,
\end{equation}
where $\sigma^2_l =   \sigma^2_x + \sigma^2_y$.  Despite its different
origin,  the  validity  condition  for   neutrino oscillations,  first
observed  by Pontecorvo in \cite{BP}, is  then quite analogous to Eq.\ 
(\ref{cond1}), {\em i.e.}
\begin{equation}
  \label{sigmal1}
\sigma_l\ \ll\ L^{\rm osc}\, .
\end{equation} 
As opposed to \cite{GK}, we do not find any new upper bound on $L^{\rm
  coh}$ within  our   solvable  model if  $\Delta   E  = 0$  and   the
requirements  given  in Eqs.\  (\ref{cond1})  and (\ref{sigmal1}) hold
true.  In fact,  in such a case neutrinos  would oscillate to infinite
distances.   Nevertheless, in realistic   situations of production and
detection of   asymptotic states, $\Delta  E$  is different from zero,
leading  to       a  physical     cut-off       for  $L^{\rm     coh}$
\cite{BP,SN/BK,KNW,LES}. For example,  for solar neutrinos, the energy
dispersion $\Delta E$ of the production system is of the same order of
the  energy thermal spread   $\Delta   E_e \sim 1$  keV   of electrons
captured by $^7Be$.

In   summary, we have  studied   neutrino oscillations in  space  in a
realistic model    in which both  the  source  and target  states have
Gaussian-form localizations.  Such a model is analytically solvable in
terms  of   complementary  error functions   within  the  perturbative
framework of field theory.  This  enables to take under close scrutiny
various approximations  used in the literature  to arrive at the known
formula (\ref{Osc})   of neutrino oscillations,  such as  the steepest
descent method or  the naive  pole-dominance approximation.   We  find
that the actual  space-dependence of particle oscillations obtained by
the solvable model ({\em c.f.}, Eqs.\ (\ref{Tamp5}), (\ref{TampS}) and
(\ref{TampF})) is very sensitive to the localization spread $\delta l$
of  the   production and   detection states.   If   $\delta l$  is  of
microscopic size such that  $\delta l^2 E \ll l$,  we then recover the
usual picture of particle oscillations  through spherical waves  given
in Eqs.\ (\ref{Osc}) and (\ref{Posc}).   If, however, the distance $l$
between the source and the detector is such  that $\delta l^2 E \gg l$
but $\delta l \ll  L^{\rm osc}$ and  $l$, neutrino oscillations  occur
through plane  waves and their apparent oscillation  length as seen by
the detector is   strongly correlated with the average  three-momentum
$\vec{p}$ of  the   initial particles   producing neutrinos.   Such  a
possibility may introduce ambiguities in the determination of the mass
differences  of neutrinos.  Which of the  two afore-mentioned forms of
propagation dominates in  different  neutrino experiments and  how can
these  be disentangled  are open challenges   that present  and future
experiments may have to face.

\bigskip\bigskip
\noindent {\bf Acknowledgments.} We  wish to thank Bill Bardeen, James
Bjorken, Stanley Brodsky, Darwin  Chang, George Gounaris, Pasha Kabir,
Wai-Yee  Keung, Mikhail  Marinov, Georg Raffelt  and Arkady Veinshtein
for discussions. Last but not least,  we are thankful to Leo Stodolsky
for illuminating conversations  concerning the stationary condition in
neutrino oscillations.  AP thanks the  theoretical groups of SLAC  and
FERMILAB for   their kind hospitality,  while  part  of the  work  was
performed.  AI  acknowledges support from  the  Alexander von Humboldt
Foundation.
 
\newpage 

\def\theequation{\Alph{section}.\arabic{equation}}
\begin{appendix}
\setcounter{equation}{0}
\section{Three-dimensional integrals}

In this   appendix we   shall  list  analytic expressions   of  useful
three-dimensional  integrals occurring in  the model under discussion.
For this purpose, we first give the formula pertaining to the Gaussian
integration
\begin{equation}
  \label{Gauss}
\int_{-\infty}^{\infty}\, d^3\vec{x}\ e^{-i\vec{a}\vec{x}}\ 
e^{-\vec{x}^2/\delta x^2}\ =\ \pi^{3/2}\, \delta x^3\, e^{-\frac{1}{4}
\delta x^2 \vec{a}^2 }\ .
\end{equation}

Another useful integral is  the  scalar propagator in three  Euclidean
dimensions
\begin{equation}
  \label{Delta}
\Delta ( \vec{l} )\ =\ \int \frac{d^3\vec{q}}{(2\pi)^3}\
    \frac{e^{i\vec{q}\vec{l}}}{q^2_j - |\vec{q}|^2 + i\varepsilon}\ .
\end{equation}
The  above  integral  can be  solved  exactly  by   virtue of Cauchy's
theorem.  The  intermediate steps of  integration may  be described as
follows:
\begin{eqnarray}
   \label{A0}
\Delta ( l ) &=& \frac{1}{4\pi^2}\ \int_0^{+\infty}
|\vec{q}|^2 d |\vec{q}|\, \int_{-1}^{+1}dz\
\frac{e^{i|\vec{q}| l z}} {q^2_j - |\vec{q}|^2 + i\varepsilon}\ =\
-\, \frac{i}{4\pi^2 l}\int_0^{+\infty}|\vec{q}|\, d |\vec{q}|\ 
\frac{e^{i|\vec{q}| l} - e^{-i|\vec{q}| l}} 
{q^2_j - |\vec{q}|^2 + i\varepsilon} \nonumber\\
 &=& -\, \frac{i}{4\pi^2 l} \int_{-\infty}^{+\infty}|\vec{q}|\, d |\vec{q}|\ 
\frac{e^{i|\vec{q}| l}} {q^2_j - |\vec{q}|^2 + i\varepsilon} \ ,
\end{eqnarray}
with $|\vec{l}| = l$.   The integrand in  the last expression  of Eq.\
(\ref{A0}) has two complex poles situated at $|\vec{q}|  = \pm q_j \pm
i\varepsilon$.  For the case $l > 0$, we may analytically continue the
contour of integration over the upper half  of the complex $|\vec{q}|$
plane.   Since the integration   over the complex semi-circle boundary
vanishes,   the   only non-vanishing   contribution   to  the integral
(\ref{A0}) comes from the pole $q_j + i\varepsilon$.  Thus, we find
\begin{equation}
  \label{A00}
\Delta ( l )\ =\ -\, \frac{i}{4\pi^2l}\ (2\pi i)\ 
\lim\limits_{|\vec{q}|\to q_j}\ (|\vec{q}|- q_j -i\varepsilon)\
\frac{|\vec{q}|\, e^{i|\vec{q}|\, l}} {q^2_j - |\vec{q}|^2 + i\varepsilon}\
=\ -\ \frac{1}{4\pi l}\ e^{iq_j l}\ .
\end{equation}

In the  following, we shall  consider the  integral  contained in Eq.\
(\ref{Tamp4}).  More explicitly, we have
\begin{eqnarray}
  \label{A1} 
{\cal M}(k,p,\vec{l},\delta y,\delta x) &=& 
\int \frac{d^3\vec{q}}{(2\pi)^3}\ \frac{e^{-\delta x^2 
(\vec{p} - \vec{q})^2/4}\ e^{-\delta y^2 (\vec{k} - \vec{q})^2/4}\
e^{i\vec{q}\vec{l}} }{q^2_j - |\vec{q}|^2 + i\varepsilon}\nonumber\\
&=& \frac{ e^{-\frac{1}{4}\delta x^2 |\vec{p}|^2 - \frac{1}{4}
\delta y^2 |\vec{k}|^2} }{2\pi^2 |\vec{L}|}\
\int_{0}^{+\infty} |\vec{q}| d |\vec{q}|\ 
\frac{e^{-\frac{1}{4} (\delta x^2 + \delta y^2 ) |\vec{q}|^2}\, 
          \sin (|\vec{q}||\vec{L}|) }{q^2_j - |\vec{q}|^2 +
i\varepsilon}\ ,
\end{eqnarray}
where $\vec{L}$ is    defined  in Eq.\  (\ref{Lvec}).   The   integral
(\ref{A1})   is generally convergent since    the integrand falls  off
exponentially in the limits  $|\vec{q}| \to \pm \infty$.  However, the
theorem by Cauchy does not apply in this case, since the integral over
the   complex  semi-circle  with   infinite   radius does not  vanish.
Therefore, we must proceed differently.

The integral we must calculate in Eq.\ (\ref{A1}) has the following
analytic form:
\begin{equation}
  \label{AI}
I(a,b,y)\ \equiv \ \frac{1}{2i}\ \int_0^{+\infty} dx\ \frac{x e^{-a^2 x^2}\ 
( e^{ix y} - e^{-ixy}) }{x^2\, +\, b^2}\ ,
\end{equation}
where $a = \frac{1}{2} \sqrt{\delta x^2 + \delta y^2} > 0$, $b = -iq_j
+ \varepsilon/(2q_j)$  ($\Re e  b > 0$),   and  $y = |\vec{L}|$  is  a
complex variable.  Instead of  considering the integral $I(a,b,y)$, we
shall first calculate
\begin{equation}
  \label{AF}
F(a,b,y)\ \equiv \ \frac{1}{2i}\ \int_0^{+\infty} dx\ \frac{x e^{-a^2 x^2}\ 
e^{ix y}}{x^2\, +\, b^2}\ ,
\end{equation}
and then obtain   $I(a,b,y)$   by the obvious  relation    
\begin{equation}
  \label{Irel}
I(a,b,y)\ =\ F(a,b,y)\, -\,  F(a,b,-y)\ . 
\end{equation}
We use    Schwinger's representation   for    the integrand   in  Eq.\
(\ref{AF}), {\em viz.}
\begin{equation}
  \label{AF1}
F(a,b,y)\ =\ \frac{1}{i}\ \lim_{R\to +\infty}\ e^{a^2 b^2}\ \int_a^{R/b}
dt\ t\ 
\int_0^{+\infty} dx\ x\, e^{- t^2 (x^2 + b^2)}\ e^{ix y}\ .
\end{equation}
In order to avoid  possible singularities originating from the complex
pole $(x^2 + b^2)^{-1}$ of the propagator, we  could perform the usual
Wick rotation and regard $x$ as a pure imaginary variable. We shall no
longer dwell upon this point, since  it also occurs  even if $a\to 0$,
in which case the result may be  read off from Eq.\ (\ref{A00}). Thus,
we can formally carry out the  Gaussian integration in (\ref{AF1}) and
rewrite $F(a,b,y)$ in the following forms:
\begin{eqnarray}
  \label{AF2} 
F(a,b,y) &=& -\, \frac{1}{2} \lim_{R\to +\infty}\ e^{a^2
b^2}\  \int_a^{R/b} dt\ t\  e^{-t^2 b^2}\ \frac{\partial}{\partial y}\
\int_{-\infty}^{+\infty} dx\ e^{- t^2  x^2 }\ e^{ix y} \nonumber\\ 
&=&
\frac{\sqrt{\pi}}{4}\  \lim_{R\to +\infty}\ e^{a^2 b^2}\ \int_a^{R/b}\
\frac{dt}{t^2}\ y\,   e^{-t^2 b^2\,  -\, y^2/(4t^2)}   \nonumber\\ 
&=&
\frac{\sqrt{\pi}}{4}\ \lim_{R\to   +\infty}\  e^{a^2  b^2}\   e^{-by}\
\int_a^{R/b}\ \frac{dt}{t^2}\ y\, e^{- [tb - y/(2t)]^2} \ .
\end{eqnarray}
Making the variable  substitution $z =  t b -  y/(2t)$ and taking  the
limit $R\to +\infty$ where appropriate, we finally arrive at
\begin{equation}
  \label{AF3} 
F(a,b,y)\ =\ \frac{\pi}{4}\ e^{a^2  b^2}\   e^{-by}\ {\rm Erfc}\Big(
ab -\frac{y}{2a}\Big)\, -\, \frac{be^{a^2 b^2}\sqrt{\pi}}{4}\  
\lim_{R\to +\infty}\  \int_a^{R/b}\ dt\ e^{-t^2 b^2\,  -\, y^2/(4t^2)}\ .
\end{equation}
In  Eq.\  (\ref{AF3}),  ${\rm  Erfc}(z)$  is   the complementary error
function of a complex argument defined as
\begin{equation}
  \label{Erfc}
{\rm Erfc}(z)\ \equiv\ 
\frac{2}{\sqrt{\pi}} \int_{z}^{+\infty} dt\ e^{-t^2}\ =\ 
1\, -\, \frac{2}{\sqrt{\pi}} \int_{0}^{z} dt\ e^{-t^2}\ .
\end{equation}
The  last  $R$-dependent  term  on  the  RHS  of  Eq.\ (\ref{AF3})  is
symmetric under the interchange  $y\to  -y$ and cancels  in $I(a,b,y)$
when the difference  in Eq.\ (\ref{Irel})  is  formed.  The so-derived
analytic result agrees with   that of Ref.\  \cite{Table} for  $y>  0$
which has been extended here to include complex  values of $y$.  After
restoring all  the  terms present  in  Eq.\ (\ref{A1}), we  eventually
obtain
\begin{eqnarray}
\label{A2}
{\cal M} (k,p,\vec{l},\delta y, \delta x) &=& - \frac{1}{8\pi |\vec{L}| }\ 
     e^{-\frac{1}{4}\,\delta x^2 (|\vec{p}|^2+q^2_j)  - 
     \frac{1}{4}\, \delta y^2 (|\vec{k}|^2 +q^2_j)}\nonumber\\
&& \hspace{-2cm} \times\, \Big[\, 
e^{iq_j |\vec{L}|}\, {\rm Erfc} \Big(-\frac{i}{2}\delta l q_j
-\frac{|\vec{L}|}{\delta l}\Big)\ -\ e^{-iq_j |\vec{L}| }\, 
{\rm Erfc} \Big( -\frac{i}{2}\delta l q_j +\frac{|\vec{L}|}{\delta l}\Big)\, 
\Big]\ ,
\end{eqnarray}
with $\delta l^2 = \delta x^2 + \delta y^2$.

It is  useful to list  a few values   of ${\rm Erfc} (z)$  for special
$z$'s as  well as derive appropriate  expansions of its argument which
have been used extensively  throughout our discussion.  In particular,
we have
\begin{equation}
  \label{Erfcval}
{\rm Erfc} (+\infty)\ =\ 0\, ,\qquad {\rm Erfc} (0)\ =\ 1\, ,\qquad
{\rm Erfc} (-\infty)\ =\ 2\, .
\end{equation}
The asymptotic behaviour of Erfc$(z)$ for $|z| \gg 1$ with $\Re z > 0$
\cite{AS} may be obtained by a series expansion, {\em viz.}
\begin{eqnarray}
  \label{Erfcexp}
{\rm Erfc}(z) &=& \frac{e^{-z^2}}{\sqrt{\pi} z}\ \Big[\, 1\ +\
\sum^\infty_{m=1}\ (-1)^m\ \frac{1\times 3\times \dots
  (2m-1)}{(2z^2)^m}\ \Big]\nonumber\\
&=& \frac{e^{-z^2}}{\sqrt{\pi} z}\ \Big[\, 1\ +\  {\cal
  O}\Big(\frac{1}{z^2} \Big)\, \Big]\, .   
\end{eqnarray}
If $\Re z < 0$, one may employ the following property:
\begin{equation}
  \label{Erfcprop}
{\rm Erfc}(z)\ = \ 2\,  -\, {\rm Erfc}(-z)\, ,
\end{equation}
and  then apply the asymptotic  expansion in Eq.\ (\ref{Erfcexp}).  By
means of Eqs.\ (\ref{Erfcexp})  and (\ref{Erfcprop}), we find  that in
the limit $\delta l \to 0$,  the expression in Eq.\ (\ref{A2}) becomes
identical to that given  in Eq.\ (\ref{A00}). Furthermore, if  $\delta
l^2  q_j\, ,\ \delta  l^2 |\vec{k}| \ll  l$ with $\vec{k}=\vec{p}$, we
obtain  the asymptotic expression  in Eq.\ (\ref{TampS1}), which is in
agreement  with    results   derived in      \cite{GS} from  different
considerations.  The latter  constitutes another  non-trivial check of
the  validity of our analytic results.   On the other hand, if $\delta
l^2 q_j\, ,\ \delta  l^2 |\vec{k}| \gg   l$ but $\delta  l \ll  l\, ,\
L^{\rm osc}$,  the transition amplitude ${\cal  M}$  has a non-trivial
dependence on the Erfc functions in Eq.\ (\ref{A2}) which multiply the
exponential  factors   $\exp    (iq_j |\vec{L}|)$ and   $\exp   (-iq_j
|\vec{L}|)$.  However, the term  depending on $\exp  (iq_j |\vec{L}|)$
leads to a  Gaussian-like  function  $\exp [-\delta l^2  (|\vec{k}|  -
q_j)^2/4   ]$ that controls momentum   conservation,  whereas the term
proportional to $\exp  (-iq_j  |\vec{L}|)$ turns out  to  be extremely
suppressed by a factor $\exp [- (l/\delta  l)^2 ]$.  As a consequence,
we find that up  to an overall  normalization constant ${\cal N}$, the
amplitude ${\cal M}$ exhibits a plane-wave behaviour in this kinematic
range (see also Eq.\ (\ref{TampS2})).

Finally,  it is amusing  to   discuss the  S-matrix limit  of  formula
(\ref{A2}).  As can be seen  from Eq.\ (\ref{Tamp4}), this corresponds
to the  case $\delta x,  \delta y \gg  l$.  For simplicity, we work in
the $\vec{k}=\vec{p}$ approximation and keep  only the leading term in
the asymptotic expansion (\ref{Erfcexp}).  After some algebra, we find
for $\vec{k}\vec{l}> 0$
\begin{eqnarray}
  \label{Smatrix}
  {\cal  M}  (\vec{k}=\vec{p},\vec{l},\delta   l) &\approx & 
-  \frac{1}{4\pi |\vec{L}|}\ 
e^{-\frac{1}{4}\,\delta l^2  (|\vec{k}|^2+q^2_j)}\,
  e^{iq_j  |\vec{L}|}\  +\  \frac{1}{4\pi^{3/2}}\, \frac{\delta    l\,
    e^{-l^2/\delta  l^2}\,  e^{i\vec{k}\vec{l}} }{l^2\, +\, \delta l^4
    (q^2_j  - |\vec{k}|^2)/4\,  -\, i\delta  l^2  \vec{k}\vec{l} }  \,
  \nonumber\\ 
&\approx  &  \frac{1}{\delta l^3  \pi^{3/2}}\ \Big[\,  -
  \frac{i\pi      e^{iq_j     \vec{k}\vec{l}/|\vec{k}|  }}{|\vec{k}|}\
  \frac{\delta  l}{2\sqrt{\pi}} e^{-\frac{1}{4}\,\delta l^2 (|\vec{k}|
    -   q_j   )^2}\nonumber\\  
&+& {\cal  P}\frac{ e^{i\vec{k}\vec{l}}
    }{q^2_j\, -\, |\vec{k}|^2  }\    +\   i\pi e^{i\vec{k}\vec{l}}
    \delta_+ (q_j^2 - |\vec{k}|^2)\ \Big]\, ,
\end{eqnarray}
where ${\cal P}$ stands for principle value.  Recognizing now the fact
that
\begin{equation}
  \label{deltaprop}
\delta (x)\ =\ \lim_{a \to \infty}\ \frac{a}{\sqrt{\pi} } \ e^{-a^2
  x^2}\ ,
\end{equation}
and noticing the   cancellation  of  the additional delta     function
$\delta_+ ( q_j^2 - |\vec{k}|^2)$  in Eq.\ (\ref{Smatrix}), it is then
not difficult to reproduce the known S-matrix result
\begin{equation}
  \label{MSmatrix}
{\cal M} (\vec{k}=\vec{p},\vec{l},\delta l\to \infty)\ =\
\frac{e^{i\vec{k}\vec{l}} }{\delta l^3 \pi^{3/2}}\ \Big[\, 
- i\pi \delta_+ ( q_j^2 - |\vec{k}|^2)\ +\ 
{\cal P}\ \frac{ 1 }{q^2_j\, -\, |\vec{k}|^2 }\, \Big]\, .
\end{equation}
In the usual S-matrix treatment, one first considers the limit $\delta
x, \delta y \to \infty$ in Eq.\ (\ref{Tamp4}) and then integrates over
the corresponding three-momentum delta functions.  Up to an irrelevant
normalization,  the result thus obtained  is identical to that of Eq.\
(\ref{MSmatrix}).

\end{appendix}

\end{document}